\documentclass[12pt]{article}
\usepackage{amsmath}
\usepackage{graphicx}
\usepackage{natbib}
\usepackage{url} 

\newcommand{\blind}{0}

\usepackage{graphicx}
\usepackage{enumerate}
\usepackage{natbib}
\usepackage{url}
\usepackage{color}
\usepackage{ulem}
\usepackage{bm}
\usepackage{multicol}
\usepackage{blindtext}
\usepackage{bbm}
\usepackage[T1]{fontenc} 
\usepackage[utf8]{inputenc} 
\usepackage[english]{babel}
\usepackage{amsmath,amssymb}
\usepackage{mathtools}

\graphicspath{{Images/}}

\usepackage{microtype} 
\usepackage{rotating} 
\usepackage{booktabs}  
\usepackage{array}     
\usepackage{multirow} 
\usepackage{enumitem}
\usepackage{algorithm}
\usepackage{algpseudocode}
\usepackage[breaklinks=true]{hyperref}  

\newcommand{\1}{\mathbf{1}}

\newcommand{\upp}{{\mathrm{upp}}}
\newcommand{\low}{{\mathrm{low}}}

\newcommand{\Bbeta}{{\boldsymbol{\beta}}}
\newcommand{\Bgamma}{{\boldsymbol{\gamma}}}
\newcommand{\Bepsilon}{{\boldsymbol{\epsilon}}}

\renewcommand{\baselinestretch}{1}

\begin{document}

\def\spacingset#1{\renewcommand{\baselinestretch}%
{#1}\small\normalsize} \spacingset{1}


\if0\blind
{
  \title{\bf Global quantile regression}
  \author{Tom\' a\v s Mrkvi\v cka\thanks{
    The authors gratefully acknowledge \textit{grant agency of Czech Republic (project number 24-11096S)}}\hspace{.2cm}\\
    Faculty of Economics, University of South Bohemia\\
    and \\
    Konstantinos Konstantinou\thanks{The authors gratefully acknowledge \textit{Swedish Research Council}}\hspace{.2cm}\\
    Department of Mathematical Sciences, \\
    Chalmers University of Technology and University of Gothenburg\\
    and \\
    Mikko Kuronen \thanks{The authors gratefully acknowledge \textit{Academy of Finland (project number 348154)}}\hspace{.2cm}\\
    Natural Resources Institute Finland (Luke) \\
    and \\
    Mari Myllym\" aki \\
    Natural Resources Institute Finland (Luke)
    }
  \maketitle
} \fi

\if1\blind
{
  \bigskip
  \bigskip
  \bigskip
  \begin{center}
    {\LARGE\bf Title}
\end{center}
  \medskip
} \fi

\bigskip
\begin{abstract}
Quantile regression is used to study effects of covariates on a particular quantile of the data distribution.
Here we are interested in the question whether a covariate has any effect on the entire data distribution, i.e., on any of the quantiles. 
To this end, we treat all the quantiles simultaneously and consider global tests for the existence of the covariate effect in the presence of nuisance covariates. 
This global quantile regression can be used as the extension of linear regression or as the extension of distribution comparison in the sense of Kolmogorov-Smirnov test.
The proposed method is based on pointwise coefficients, permutations and global envelope tests. The global envelope test serves as the multiple test adjustment procedure under the control of the family-wise error rate and provides the graphical interpretation which automatically shows the quantiles or the levels of categorical covariate responsible for the rejection. 
The Freedman-Lane permutation strategy showed liberality of the test for extreme quantiles, therefore we propose four alternatives that work well even for extreme quantiles and are suitable in different conditions.
We present a simulation study to inspect the performance of these strategies, and we apply the chosen strategies to two data examples.
\end{abstract}

\noindent%
{\it Keywords:}  Distribution comparison, Global envelope test, Multiple comparison problem, Permutation test, Significance testing, Simultaneous testing 

{\it Disclosure statement:} 
The authors have no competing interests to declare.
\vfill

\newpage
\spacingset{1.5} 
\section{Introduction}\label{sec:intro}

Quantile regression is used in many research fields to model the quantiles or full conditional distribution of the response variable rather than the mean and variance when assumptions of the ordinary linear model do not hold.
If effects of covariates are tested simultaneously for all quantiles, the problem of multiple testing arises.
We refer to the treatment of all quantiles under the control of family-wise error rate as global inference for quantile regression. This can be used to infer if a covariate influences the response variable in any of the quantiles. As a special case, the global inference can be used to compare distributions with or without additional covariates. 

In quantile regression the user is often interested in estimation of the effect of a certain covariate together with its confidence interval plotted simultaneously for all quantiles. The confidence bands help to understand the analysis results.
Usually these confidence intervals are however computed pointwisely, i.e., for every quantile independently. There are several choices of methods for pointwise estimation of confidence intervals as summarized, e.g., by \citet{Koenker2005} and implemented in the R package quantreg \citep{quantreg}, with visualization. 

In this paper, we are interested in making inferences for all quantiles simultaneously, along with a graphical interpretation that could be used to determine for which quantiles the effect of a covariate is present. 
This global inference can be viewed as an extension of quantile regression, which tests the effect of the covariate locally at a specific quantile, whereas the proposed inference tests the effect globally for all quantiles. It can also be viewed as a direct and valid extension of inference in linear regression, in the sense that the effect of a covariate on the entire distribution is tested rather than just its effect on the mean. Further, in the case of a categorical covariate, it can be viewed as an extension of distribution comparison in the sense of the Kolmogorov-Smirnov test, allowing for nuisance covariates.
We propose a global test for the significance of the effect of a covariate in a quantile regression. 

The problem of simultaneous inference for quantile regression can be solved by testing the effect of a covariate for all quantiles pointwisely by methods reviewed, e.g., in \citet{koenker1994confidence}
and applying a multiple correction method, e.g., the Holm-Bonferroni correction \citep{Holm1979} in order to solve the multiple testing problem.
Also, recently, new methods for simultaneous confidence bands were developed analytically. For example \citet{BelloniEtAl2014} and \citet[][Chapter 15]{KoenkerEtAl2018} discuss the simultaneous confidence bands for a quantile process $\Bbeta(\tau)$ on $[0,1]$
based on asymptotic theory. These bands are however valid only under complex regularity conditions. On the other hand, \citet{PengFine2009} proposed a cumulative approach in order to summarize the covariates effect of all quantiles in one number. This can be used to deduce if the effect is globally significant, but it can not be used to infer which quantiles are significant. Another global problem was considered in \citet{Khmaladze1982} and \citet{KoenkerXiao2002}, namely the constantness of the effect of all covariates.

In order to achieve  global inference for quantile regression,
we rely on permutation methods in this paper. \citet{CadeJon2006} used the Freedman-Lane (FL) permutation strategy \citep{FreedmanLane1983} for the quantile regression. This strategy is regarded as the most precise method in testing a covariate effect of a univariate or functional linear models in the presence of nuisance covariates \citep{AndersonRobinson2001, AndersonTerBraak2003, WinklerEtal2014}. \citet{CadeJon2006} also proposed an improvement of the FL procedure for quantile regression, which we also investigate in this paper. They used it for testing with a univariate test statistic which reflects the location or scale of the distribution only.
\citet{DitzhausEtAl2021} proposed to use permutations for quantile regression, too, but they proposed only simple permutation of the data, i.e.\ the strategy of the one-way ANOVA problem (even thought this was applied for factorial design of two-way ANOVA). Similarly as \citet{CadeJon2006}, they concentrated on univariate test statistics such as the median or interquartile distance. 

Here, we are interested in testing the effect for all quantiles
at the global significance level $\alpha$. 
We investigate the suitability of various permutation strategies for the given aim. 
It turns out that the FL permutation strategy does not perform well for global quantile regression, due to its liberality for extreme quantiles. Therefore, we propose several alternative permutation strategies, which perform better for extreme quantiles.

Our method for solving the problem of multiple testing is based on global envelope tests \citep{MyllymakiEtal2017,  MrkvickaEtAl2022, MyllymakiMrkvicka2020} recently developed for spatial statistics and functional data analysis. This method allows 
to use a functional test statistic 
and have the global significance level $\alpha$.
Besides, it allows us to draw the $100(1-\alpha)\%$ global envelope that represents the acceptance region under the null model of no effect of a certain covariate under the presence of other covariates. If the observed effect of the covariate is not fully contained in the global envelope, the test is significant at the global significance level $\alpha$. 
Further, the test shows the quantiles which are the reason for a potential rejection of the null hypothesis, suggesting how the covariate affects the distribution of the response variable.

Since the global envelope test is based on ranks, it has no assumptions on the distribution of the functional test statistics, neither the homogeneity of the distribution of the test statistic along its domain. The only assumption is exchangeability of the test statistic under the permutation strategy. Some of the studied permutation strategies fullfil the exchangeability but some do not. For instance, the famous FL permutation strategy does not satisfy the exhangeability in presence of nuisance covariates. Therefore, we study via simulation study which of the alternative permutation strategies match the best the preset significance level and have the highest power.

Due to the nonparametric nature of the global envelope test, we can test continuous, categorical effect, interactions, and within the categorical effect also the differences between the groups via the joined functional test statistics \citep{MrkvickaEtal2017}.

The 
rest of the paper is organized as follows. Section \ref{sec:background} gives the necessary background on quantile regression and global envelope tests. Section \ref{sec:globalquantileregression} explains the proposed global test. Section \ref{sec:permstrategies} describes the different permutation strategies to generate simulations under the null model of no effect of the interesting covariate. The performance of the permutation strategies together with the global test are then investigated in Section \ref{sec:simstudy}.
Section \ref{sec:dataexamples} applies the chosen tests to analyse two data sets.
Section \ref{sec:discussion} is for discussion of the results and extensions.
The implementation of the proposed method will be available in the R package GET \citep{MyllymakiMrkvicka2020}.

\section{Notation and background}\label{sec:background}

\subsection{Linear quantile regression}

Classical linear regression models focus on modelling the conditional expectation of a response variable $\mathbf{Y}$ given a set of covariates $\mathbf{X}$. In linear regression, the mean response is modelled as a linear combination of the regression parameters $\Bbeta$ and the covariates $\mathbf{X}$, \ i.e. $\mathbb{E}(\mathbf{Y}\mid \mathbf{X})= \mathbf{X}\Bbeta$, and estimation of the regression coefficients is performed by minimizing the sum of squared residuals. However, linear regression models are often insufficient either due to violations of the linear model's assumptions or due to the interest being in the tails of the distribution rather than its mean. Hence, analysis of covariate effects across the conditional distribution of the response variable requires more flexible statistical modeling than traditional linear regression only.

Quantile regression introduced by \cite{koenker1978regression} focuses on the modelling of the conditional quantiles of the response variable. That is, for any $\tau\in[0,1]$, the $\tau-$quantile of the conditional distribution of the response $Y_i$ given a set of covariates 
{$\mathbf{X}_i$, 
\begin{equation}\label{eq:Qitau}
    Q_{Y_i|\mathbf{X}_i}(\tau) = \inf\{y: F_{Y_i|\mathbf{X}_i}(y)\geq \tau\}=\mathbf{X}_i^T\Bbeta(\tau),\, i = 1, \ldots , n
\end{equation}
where $F_{Y_i|\mathbf{X}_i}$ is the conditional cumulative distribution function of $Y_i$ given $\mathbf{X}_i$, and $\Bbeta(\tau)$ is the regression coefficient of the model for the $\tau$-quantile. For instance, the quantile regression for $\tau=0.5$ defines the linear model for the conditional median, a robust alternative to the standard linear model.

Unlike classical linear regression, which has a closed formula for the estimator of the regression coefficients, estimating the parameters of quantile regression requires solving an optimization problem. The regression coefficients $\Bbeta(\tau)$ are estimated by minimizing the weighted absolute residuals 
\begin{equation}
\hat{\Bbeta}(\tau)= \min_{\Bbeta\in\mathbb{R}^d}\sum_{i=1}^n \rho_{\tau}(
{Y_i}-\mathbf{X}_i^T \Bbeta)
\label{eq:optim}
\end{equation} 
where $\rho_\tau(u) = u\left(\tau - \mathbbm{1}(u<0)\right)$, i.e., $\rho_\tau(u) = u\tau$ if $u\geq 0$ and $-u(1-\tau)$ if $u<0$. The optimization problem in Equation \eqref{eq:optim} can efficiently be solved by linear programming methods \citep{dantzig2016linear,portnoy1997gaussian}.
We used the R library quantreg \citep{quantreg} for the estimation of $\hat{\Bbeta}(\tau)$.

\subsection{Inference for quantile regression}

Studying the effect of the covariates of interest on quantiles of 
the conditional distribution of the response, requires inference of the quantile regression process $\Bbeta(\tau)$ on $[0,1]$. In the literature there exist three main approaches to construct confidence intervals for $\hat{\Bbeta}(\tau)$. 

The first approach assumes that under some mild conditions, the estimated regression quantiles are asymptotically normal \citep{Koenker2005}. Calculating the standard error requires the estimation of the so called sparsity function $s(\tau) = [f(F^{-1}(\tau))]^{-1}=\frac{d}{d\tau}F^{-1}(\tau)$, where $f$ is a probability density function such that $f=F'$. The sparsity function $s(\tau)$ can be estimated by $\hat{s}_n(\tau) = [\hat{F}_n^{-1}(\tau+h_n)-\hat{F}_n^{-1}(\tau-h_n)]/ 2h_n$ where $h_n$ is a bandwidth which tends to zero as $n\to\infty$ and needs to be selected, and $\hat{F}_n$ is the empirical cumulative distribution function, or by kernel smoothing. 
The estimator $\hat{s}_n(\tau)$ is unstable when the assumption that the errors are iid is violated. In the case with non iid errors, a Huber estimate of the limiting covariance matrix needs to be computed \citep{koenker1994confidence}. This case can be treated by assuming that $Q_{Y_i|\mathbf{X}_i}(\tau)$ is locally linear in $\mathbf{X}_i$ \citep{koenker1999goodness}.
 For the remainder of the paper we refer to this method as the ``NID'' method. The quantreg package recommends using the NID method for data with more than 1000 datapoints {as this method is very fast \citep{chen2005computational}}. On the contrary, this method is not ideal for small samples, as the methods for automatic bandwidth selection, for instance the method in  \cite{hall1988distribution}, tend to give large bandwidths, which often result in violations of the local linearity assumption.

The second class of methods are the rank-score methods which construct the confidence intervals by the inversion of the rank-score test \citep{GutenbrunnerEtAl1993,koenker1994confidence,koenker1999goodness}. The rank-score methods, avoid the estimation of the sparsity function and are more robust to model assumptions. However, those methods require solving a parametric linear programming problem. {Therefore, this approach is slow for large samples as its computational complexity is exponential in $p$ and $n$ \citep{chen2005computational,kocherginsky2005practical}}. In the quantreg package, the rank method is used by default for small samples ($n< 1000$).

The third method for constructing confidence intervals is based on resampling strategies \citep{efron1979bootstrap}. {Most common methods are based on bootstrapping the the pairs of the response and explanatory variables \citep{hahn1995bootstrapping} or bootstrapping the residuals \citep{bickel1981asymptotic}. In the residual bootstrap exchangeability of the residuals needs to be assumed. Recently, there have been a lot of research using bootstrap techniques for estimating standard errors in the quantile regression setting \citep{parzen1994resampling,he2002markov,kleiner2014scalable}}.

However, all the methods above concern local inference, but we are interested in simultaneous inference for $\Bbeta(\tau)$, $\tau\in[0,1]$. In this paper, we propose an inference method using permutation based global envelopes test. The proposed test is compared with the Holm-Bonferroni adjusted local NID test (see above). 

Another important question in quantile regression is if the effect of all covariates can be considered constant for all quantiles. It was studied in \cite{KoenkerXiao2002}. They proposed tests for the hypothesis that a linear model specification is of the location shift (i.e. the effect of covariates for all quantiles is constant) or location-scale shift form (i.e. the  covariates affect only mean and variance of the response distribution). The tests are based on the approach proposed by \cite{Khmaladze1982}.

\subsection{Quantile regression for modelling distributions}

There are tests to test differences between the two distributions. The two-sample Kolmogorov-Smirnov  test is maybe the most well known.
Here we only remark that the global quantile regression with a categorical predictor can also be used to solve the problem of finding the differences between the distributions (two or more), not only when the categorical predictor is the only covariate of the model but also in the presence of further nuisance covariates. The proposed global quantile regression can determine not only if there is a difference, but it can also determine for which $\tau$'s the difference is significant at the global significance level.

\subsection{Global envelope tests}
Global envelope tests are non-parametric Monte-Carlo tests for multivariate or functional summary statistics \citep{MyllymakiEtal2017}. Let $\mathcal{T}=(\tau_1,\dots, \tau_d)$ be {the vector of} $d$ discrete values where the statistic is evaluated. Further, let $\mathbf{T}_0 = (T_{01}, \dots, T_{0d}) = (T_{0}(\tau_1), \dots, T_{0}(\tau_d))$ stand for the $d$-dimensional discretization of the empirical statistic and $\mathbf{T}_1,..., \mathbf{T}_{s}$ be the corresponding statistics for $s$ data sets simulated under the ``null model''.  The tests are global in the sense that the test is performed simultaneously for all $\tau\in\mathcal{T}$, i.e. the family-wise error rate is controlled by the {
prespecified} significance level $\alpha$. The advantage of global envelope test is that it allows for graphical interpretation of the test result by a global envelope that represents the acceptance region of the test:
A $100(1-\alpha)\%$ global envelope is a band $(\mathbf{T}_{\low}^\alpha,\mathbf{T}_{\upp}^\alpha)$ with $\mathbf{T}_{\low}^\alpha = (T_{\low, 1}^\alpha, \dots, T_{\low, d}^\alpha)$ and $\mathbf{T}_{\upp}^\alpha = (T_{\upp, 1}^\alpha, \dots, T_{\upp, d}^\alpha)$, constructed under the null model, such that the probability that $\mathbf{T}_0$ is completely within the envelope is equal to $1-\alpha$.
Therefore, the empirical test statistic $\mathbf{T}_0$ goes outside the given $100(1-\alpha)$\% global envelope for some $\tau$  if and only if the global test rejects the null hypothesis {($p<0.05$)}. 
The $\tau$'s where $\mathbf{T}_0$ goes outside the envelope are responsible for the rejection of the test. 

Global envelopes are constructed by ranking the statistics $\mathbf{T}_0,\dots,\mathbf{T}_{s}$ based on a ranking measure $E$. The ranking is then used to identify the $\alpha{(s+1)}$ most extreme vectors. Examples of the ranking measures, which allows for one-to-one correspondence between formal and graphical results, are the extreme rank length measure \citep{NarisettyNair2016,MyllymakiEtal2017}, the continuous rank measure \citep{Hahn2015} and the area measure \citep{MrkvickaEtAl2022}. For a more rigorous description of the available ranking measures you are referred to \cite{MyllymakiMrkvicka2020} and references therein. Now, let $E_i<E_j$ be interpreted as $\mathbf{T}_i{(\tau)}$ is more extreme than $\mathbf{T}_j{(\tau)}$ and let $E_{(\alpha)}\in\mathbb{R}$ be the largest $E_i$ such that
\begin{equation*}
    \sum_{i={0}}^s \1(E_i<E_{(\alpha)})\leq \alpha (s+1)
    \label{eq:ea}
\end{equation*} 
and let  $I_{(\alpha)}$ denote the set of vectors less than or as extreme as $E_{(\alpha)}$. Then, a {$100(1-\alpha)$\%} global envelope based on the measure $E$ is given by \begin{equation*}
    \left(T^{(\alpha)}_{\low \ k},T^{(\alpha)}_{\upp \ k}\right) =
\left(\min_{i\in I_{(\alpha)}}T_{ik},\max_{i\in I_{(\alpha)}}T_{ik}\right)\quad \text{for $k=1,...,d$}.
\label{envelope}
\end{equation*}

The validity of global envelope tests is independent of the distribution or potential inhomogeneity of the distribution of the test statistic along its domain. However, in order for the global envelopes to achieve desired type I errors, {
the test statistics $\mathbf{T}_0,...,\mathbf{T}_{s}$ must be exchangeable}. The exchangeability depends on the permutation strategy used to obtain the replications of the test statistic under the null model. 

Any functional measure $E$ can be used to rank the statistics $\mathbf{T}_0,...,\mathbf{T}_{s}$, but only those which satisfies the one to one correspondence between formal results and their graphical interpretation represented by the global envelope are considered in this work.

\section{Global quantile regression}\label{sec:globalquantileregression}

Assume the quantile regression model  
\begin{equation}\label{eq:qr_full}
 Q_{\mathbf{Y}|\mathbf{X},\mathbf{Z}}(\tau) = \mathbf{X}\Bbeta(\tau) + \mathbf{Z}\Bgamma(\tau) \text{ for all } \tau\in\mathcal{T},
\end{equation}
where $Q_{\mathbf{Y}|\mathbf{X},\mathbf{Z}}(\tau) = (Q_{Y_1|\mathbf{X}_1,\mathbf{Z}_1}(\tau), \dots, Q_{Y_n|\mathbf{X}_n,\mathbf{Z}_n}(\tau))$ is a $n\times 1$ vector of conditional $\tau$-quantiles of $Y_1,\dots,Y_n$, $\mathbf{X}$ is a $n\times p$ matrix of the interesting covariates, $\mathbf{Z}$ is a $n\times q$ matrix of nuisance covariates, 
$\Bbeta(\tau) = (\beta_1(\tau), \dots, \beta_p(\tau))$ and $\Bgamma(\tau) = (\gamma_1(\tau), \dots, \gamma_q(\tau))$ are the corresponding parameter vectors of dimensions $p\times 1$ and $q\times 1$, respectively, and
$\mathcal{T} = \{\tau_1,\ldots , \tau_d$\} is a discrete set of quantiles we are interested in. The null hypothesis of interest is
\begin{equation}\label{eq:H0}
H_0: \beta_j(\tau)=0 \text{ for all } j=1,\dots,p \text{ and } \tau\in \mathcal{T}.    
\end{equation}
Our aim is to construct a test with the family-wise error rate control for all $\beta_j$, $j=1,\dots,p$ and $\tau \in \mathcal{T}$, i.e., global quantile regression test of significance of covariates contained in $\mathbf{X}$. We propose the following strategy for this purpose:

\begin{algorithm}
\caption{Global inference for quantile regression \eqref{eq:qr_full} using permutation schemes}
\label{alg:QRtest}
    \begin{enumerate}
    \item For observed data, compute the test vector
    \begin{equation}\label{eq:T0}
    \mathbf{T}_0 = 
    (\beta_1(\tau_1), \dots, \beta_1(\tau_d), \ldots \beta_p(\tau_1), \dots, \beta_p(\tau_d)) 
    \end{equation}
    containing all the coefficients of the vectors $\Bbeta(\tau_1), \dots, \Bbeta(\tau_d)$, rearranged for better visualization.
    \item Simulate $s$ replicates of data under the null hypothesis \eqref{eq:H0}.
    \item Compute the test vectors for the $s$ simulated data, and obtain $\mathbf{T}_1,\dots,\mathbf{T}_s$.
    \item Apply a global envelope test to $\mathbf{T}_0,\mathbf{T}_1,\dots,\mathbf{T}_s$.
    \end{enumerate}
\end{algorithm}

Global envelope testing provides a global $p$-value, the graphical interpretation that determines the $\tau$'s and the elements of the vector $\Bbeta$ that are responsible for the rejection in the global test (see the data study examples for detailed description of graphical interpretation). 
Since we observe all parameters in $\mathbf{T}_0$, we perform - simultaneously with the global test - a post-hoc test in cases when the covariate is categorical. This means that all levels of the categorical covariate are tested to have different effect than the reference level. 
{The generation of the data under the null hypothesis \eqref{eq:H0} is a critical part of the test; in the following section, we will describe different alternatives for this purpose.}

Remark here that the global envelope test produces the {acceptance and} rejection regions for the global null hypothesis, whereas usually the pointwise confidence intervals for the parameters of the model are obtained in quantile regression procedures.

\section{Permutation strategies for quantile regression}\label{sec:permstrategies}

In the following, we introduce six permutation strategies  
as candidates for producing simulations under the null hypothesis \eqref{eq:H0}.
We note that exhangeability of the test statistics $\mathbf{T}_0,\mathbf{T}_1,\dots,\mathbf{T}_s$ is satisfied only for the permutation strategy for categorical covariates described in Section \ref{sec:cat_permstrategy}.

\subsection{Freedman-Lane (FL)}\label{sec:FL}

Several approximative permutation methods have been proposed to test the significance of one or more regression coefficients in univariate and functional linear regression models for conditional means. Freedman-Lane procedure \citep{FreedmanLane1983} has been found to be the method that is closest to being exact, i.e., reaching the nominal significance level \citep{AndersonRobinson2001, AndersonTerBraak2003}. In the following, we explain how the replicates of data under the null hypothesis \eqref{eq:H0} are obtained in the Freedman-Lane permutation scheme. The general idea of the method is to permute the residuals of the reduced model which does not contain the interesting covariates.

New data $\mathbf{Y}^*$ are generated by the following steps:
\begin{enumerate}
\item Fit the reduced model
\begin{equation}\label{eq:qr_reduced}
 Q_{\mathbf{Y}|\mathbf{Z}}(\tau) =  \mathbf{Z}\Bgamma(\tau) \text{ for all } \tau\in\mathcal{T}
\end{equation}
to obtain the estimated coefficients $\widehat{\Bgamma}(\tau)$.
\item Compute the residuals 
\begin{equation}\label{eq:qr_reduced_residuals}
\epsilon_{i}(\tau) = Y_i(\tau) - \textbf{Z}_i^T\widehat{\Bgamma}(\tau)
\end{equation}
of the model \eqref{eq:qr_reduced} for $i=1,\dots,n$ and $\tau\in\mathcal{T}$.
\item Permute the rows of the $n \times d$ residual matrix $\Bepsilon$ to produce the permuted residual matrix $\Bepsilon^*$.
\item Construct the permuted data 
    \begin{equation}\label{eq:permuteddata}
    \mathbf{Y}^*(\tau) = \textbf{Z}\widehat{\Bgamma}(\tau)+\Bepsilon^*(\tau) \text{ for every } \tau \in \mathcal{T},
    \end{equation}
  where $\Bepsilon^*(\tau)$ correspond to columns of $\Bepsilon^*$.
\end{enumerate}

\subsection{Freedman-Lane with removal of zero residuals (FL+)}\label{sec:FL+}

\citet{CadeJon2006} suggested an 
enhancement to the permutation strategy of \citet{FreedmanLane1983} in the case of quantile regression.
Their adjustment excludes from the permutations the zero residuals that are inherent in the quantile regression. 
That is, in the step 4.\ of the Freedman-Lane simulation (see Section \ref{sec:FL}), for every $\tau$, new permuted data $\mathbf{Y}^{**}(\tau)$ 
are constructed from the $\mathbf{Y}^*(\tau)$ of Equation \eqref{eq:permuteddata} in the Freedman-Lane permutation by removing $q-1$ elements corresponding to zero residuals.
The new data $\mathbf{Y}^{**}(\tau)$ will have only $n-q+1$ observations.

\subsection{Within categorical nuisance (WN)}\label{sec:cat_permstrategy}
\label{perm1}

In the case that the quantile regression model \eqref{eq:qr_full} includes only categorical nuisance covariates, it is possible to employ simple permutations of the response variable within each level of the categorical nuisance covariates:
Assume that there is a categorical nuisance covariate which has $K$ levels. If there are more than one categorical nuisance covariates, every group of the first nuisance covariate can be decomposed into smaller groups according to the second nuisance covariate, etc. The decomposition then forms a new categorical covariate, say, with $K$ levels. Because of the decomposition, the interactions of the nuisance factors are always present in the permutations. New data are in this case generated as follows:
\begin{enumerate}
    \item Split the data into subsets based on the $K$ levels of $\mathbf{Z}$. Let $(\mathbf{Y}^{(k)},\mathbf{X}^{(k)},\mathbf{Z}^{(k)})$, with $k=1,\dots,K$, be the $K$ subsets.
    \item Within each subset $k=1,\dots,K$,
    permute the elements of each $\mathbf{Y}^{(k)}$ to produce $\mathbf{Y}^*{(k)}$ and consequently $\mathbf{Y}^*$. 
\end{enumerate}

\subsection{Simple permutation with removal of the location effect of the nuisance covariates  (RL)}\label{sec:cont_permstrategy}

In this permutation scheme, the mean effect of nuisance covariates is removed using a linear model and residuals of the fitted model are then permuted to simulate under the null hypothesis. We adjust Algorithm \ref{alg:QRtest} for this procedure as specified in Algorithm \ref{alg:RLtest}.

\begin{algorithm}
    \caption{Global inference for quantile regression \eqref{eq:qr_full} with removal of the location effect of the nuisance covariates (RL).}\label{alg:RLtest}
\begin{enumerate}
   \item Fit the mean linear model 
$$
\mathbf{Y} = \mathbf{Z}\Bgamma + \Bepsilon_Z.
$$

    \item Fit
the quantile regression model for the residuals of the linear model from 1., 
\begin{equation}\label{eq:qr_epsilonz}
Q_{\Bepsilon_{Z} \mid \mathbf{X}}(\tau) = \mathbf{X}\Bbeta(\tau) \text{ for all } \tau\in\mathcal{T}.
\end{equation}
The test vector $\mathbf{T}_0$ is specified according to Formula \eqref{eq:T0} from the estimated coefficients of the model \eqref{eq:qr_epsilonz}.

    \item Permute the residuals $\Bepsilon_Z$ to obtain simulated data $\Bepsilon_Z^*$. Repeat this $s$ times.
    \item Compute the test vectors for the $s$ simulated data, and obtain $\mathbf{T}_1,\dots,\mathbf{T}_s$.
    \item Apply a global envelope test to $\mathbf{T}_0,\mathbf{T}_1,\dots,\mathbf{T}_s$.

\end{enumerate}
\end{algorithm}

Note here that due to the specificity of the quantile regression, it is necessary to include always the intercept in between the interesting covariates in $X$ in steps 2.\ and 4.\ of this algorithm. This holds also for the next two algorithms (Sections \ref{sec:cont_permstrategy2} and \ref{sec:quantile_permstrategy3}).

\subsection{Simple permutation with removal of the location and scale effect of the nuisance covariates (RLS)}\label{sec:cont_permstrategy2}

In this permutation scheme, the scaling of the residuals is added to Algorithm \ref{alg:RLtest} in order to remove the scale of the nuisance effect.
That is, the permutation scheme is as in the Algorithm \ref{alg:RLtest} with changing of step 1.\ with
\begin{itemize}
    \item[1'] Fit the mean linear model 
    $\mathbf{Y} = \mathbf{Z}\Bgamma + \Bepsilon'_Z$,
    then fit the mean linear model
    $\text{abs}(\Bepsilon'_{Z} )=\mathbf{Z}\boldsymbol{\omega}+\Bepsilon^{''}_{Z}$.
    Set $\Bepsilon_{Z} = \Bepsilon'_{Z}/(\mathbf{Z}\boldsymbol{\omega})$.
\end{itemize}

\subsection{Simple permutation with removal of the quantile effect of the nuisance covariates (RQ)}

\label{sec:quantile_permstrategy3}

In this permutation scheme, effects of nuisance covariates are removed using a quantile regression model and residuals of the fitted model are then permuted to simulate under the null hypothesis \eqref{eq:H0}.
The permutation scheme is as in Algorithm \ref{alg:RLtest} with changing of steps 1.\ and 2.\ with

\begin{itemize}
    \item[1''] Fit the quantile regression model
$$
Q_{ \mathbf{Y} | \mathbf{Z}}(\tau) = \mathbf{Z}\Bgamma(\tau)  \text{ for all } \tau\in\mathcal{T},
$$ 
from where the residuals $\Bepsilon_Z=(\Bepsilon_Z(\tau_1), \ldots , \Bepsilon_Z(\tau_d))$ are obtained.
    \item[2''] Consider
$d$ quantile regression models for the residuals $\Bepsilon_Z(\tau_1), \ldots , \Bepsilon_Z(\tau_d)$, 
\begin{equation}\label{eq:qr_epsilonz_RQ}
Q_{\Bepsilon_{Z}(\tau_1) | \mathbf{X}}(\tau_1) = \mathbf{X}\Bbeta(\tau_1), \ldots , Q_{\Bepsilon_{Z}(\tau_d) | \mathbf{X}}(\tau_d) = \mathbf{X}\Bbeta(\tau_d).
\end{equation}
Compute the test vector $\mathbf{T}_0$ according to Formula \eqref{eq:T0} from the estimated coefficients of the models \eqref{eq:qr_epsilonz_RQ}.
    
\end{itemize}

In this permutation scheme, similarly like in the FL+ scheme, the different data are used for different $\tau$'s, but the permutations are kept the same.

\section{Simulation study}\label{sec:simstudy}

We assumed the quantile regression model \eqref{eq:qr_full} and studied
the performance of the global test for the hypothesis \eqref{eq:H0} under different permutation schemes (see Table \ref{table:1}). 
The performance was investigated in terms of power and type I errors.
Additionally, the permutation based methods were also compared with Holm-Bonferoni corrected $p$-values obtained using the NID method as implemented in the quantreg package as well as the minimum pointwise $p$-value without any correction. 

In each experiment, the interesting covariate $X$ influences the distribution of the response variable $Y$. In addition, the nuisance covariates $Z$ and $Z_1$ affect the response distribution. We considered three different nuisance effects, namely location shift, location-scale shift, and shape shift effects. To investigate the validity of the permutation strategies in case of model misspecification, we designed scenarios where the underlying assumptions of the permutation strategy is not met. For instance, using a permutation strategy based on the nuisance location shift assumption, when the nuisance affects the shape of the response distribution. Furthermore, we studied how correlation between $X$ and $Z$ affects the performance of the methods.

Our observations consist of realizations of $X, Z, Z_1$ and $Y$ from their corresponding distributions. In all tests below, unless otherwise specified, we used the following choices:
\begin{itemize}
    \item All the global envelope tests (first six tests of Table \ref{table:1}) were based on 1000 permutations. 
    \item We considered 10 equally spaced quantiles $\tau$ varying from 0.01 up to 0.99, except for FL+ where we also considered 10 quantiles $\tau$ on the interval from 0.1 to 0.9. The tests that did not consider the extreme quantiles $\tau$ are denoted by an asterisk(*) in the figures. 
\end{itemize}

We performed the first set of experiments as in Section \ref{sec:simstudy_tails} also with 100 equally spaced quantiles $\tau$ varying from 0.01 to 0.99. 
The results were correspondent to those with 10 $\tau$ values
with respect to their significance level,
except for the PH procedure. (The NC method was not included to the experiment.) 
The PH procedure had lower empirical significance levels with 100 $\tau$ values than with 10 $\tau$ values: it was conservative in the cases where it was exact for 10 $\tau$ values, but it persisted to be liberal in cases where it was liberal for 10 $\tau$ values.
Therefore and for the reason of faster computing time, we present below the results only for the case of 10 $\tau$ values as specified above.

\begin{table}[h]

\caption{Description and abbreviations of the tests investigated in the simulation study. The first six methods are based on global quantile regression (GQR) with different permutation strategies.}

\scalebox{0.9}{
\begin{tabular}{l| l}
\label{table:1}
 Test description &  Abbreviation\\
 \hline
 GQR using the Freedman-Lane permutation&  FL \\
 GQR using the extension of the Freedman-Lane permutation  &  FL+\\
 GQR using the permutation that removes the location nuisance effect &  RL\\
 GQR using the permutation that removes the location-scale nuisance effect &  RLS\\

 GQR using the permutations for categorical nuisance&  WN\\
GQR using the permutation that removes the quantile
nuisance effect &RQ\\
Pointwise $p$-values adjusted using Holm-Bonferroni method &  PH\\
Minimum pointwise $p$-value &NC
\end{tabular}
}
\end{table}

\subsection{Sensitivity to differences in the tails of the distributions}\label{sec:simstudy_tails}

In the first two simulation experiments,  $X$ was categorical with two levels and the two distributions corresponding to the levels of $X$ differed in the tails. For the nuisance covariate, we considered different alternatives. It was either categorical or continuous.
In Experiment \eqref{Exp:1}, it affected either the location or location and scale of the response distribution, while in Experiment \eqref{Exp:2} we considered a "noise" nuisance covariate affecting the shape of the response distribution. 
More precisely, in Experiment \eqref{Exp:1}, 
  \begin{equation}\tag{I}
 \qquad    
\left\{
	\begin{array}{ll}
 \label{Exp:1}
		X \sim \text{Bernoulli}(0.5)\\
     Y' \mid X \sim \left\{
	\begin{array}{ll}
 N(0,1) & \text{if } X=0\\
t_4 & \text{if } X=1
 \end{array}
 \right.\\
   Z \sim F_Z\\
  Y =(1+aZ)Y'+bZ
	\end{array}
\right.  \end{equation} 
where $a,b \in \mathbb{R}$ and $F_Z$ is the distribution of the nuisance variable for which we considered the following four alternatives: 
\begin{itemize}
    \item[(Ia)] Continuous $Z$ with effect on the location, $F_Z = \text{Unif}(0,1.5)$, $a=0$, $b=1$
    \item[(Ib)] Continuous $Z$ with effect on the location and the scale, $F_Z = \text{Unif}(0,1.5)$, $a=1$, $b=1$
    \item[(Ic)] Categorical $Z$ with effect on the location, $F_Z = \text{Bernoulli}(0.5)$, $a=0$, $b=0.1$
    \item[(Id)] Categorical $Z$ with effect on the location and the scale, $F_Z = \text{Bernoulli}(0.5)$, $a=0.1$, $b=0.1$
\end{itemize}

In Experiment \eqref{Exp:2}, 

\[ 
\left\{
	\begin{array}{ll}\tag{II}\label{Exp:2}X \sim \text{Bernoulli}(0.5)\\
 Y' \mid X \sim \left\{
	\begin{array}{ll}
 N(0,1) & \text{if } X=0\\
t_4 & \text{if } X=1
 \end{array}
 \right.\\
   Z \sim F_{Z}\ \\
   Z_1 \sim \text{Unif}(0,1.5)\\
		 \epsilon \sim N(1,0.04)\\
   Y = \left\{
	\begin{array}{ll}
 \epsilon & \text{if } Z < Z_1
 \\
   Y' & \text{otherwise}
 \end{array}
 \right.
	\end{array}
\right.\]
where both $Z$ and $Z_1$ are nuisance covariates and $F_Z$ is the distribution of the nuisance covariate $Z$ with the following two alternatives: 

\begin{itemize}
    \item[(IIa)] Continuous $Z$ with $F_Z=\text{Unif}(0,1)$,
    \item[(IIb)] Categorical $Z$ with $F_Z=\text{Bernoulli}(0.5)$.
\end{itemize} 

For all cases of Experiments \eqref{Exp:1} and \eqref{Exp:2}, we simulated two data sets with $M=100000$ datapoints, one for testing the empirical significance level ($D_{\text{sign}}$) and one for testing for power of the tests ($D_{\text{power}}$). For both datasets, we first simulated $M$ realisations of the interesting covariate $X$ from the Bernoulli(0.5) distribution and $M$ realisations of the nuisance covariate $Z$ from $F_Z$. 
For $D_{\text{power}}$, we then simulated the response variable $Y$ as specified above.
For $D_{\text{sign}}$, the only difference in the construction was that the values of $Y'$ of Experiments \eqref{Exp:1} and \eqref{Exp:2} were simulated from $N(0,1)$, both for $X=0$ and $X=1$, making the two distributions to coincide.
We then used simple random sampling without replacement to obtain samples of size $N = 10, 50, 100, 200, 300, 500, 800, 1000$. For each sample size $N$, we drew 1000 independent samples.
For each sample of data, we then performed the tests of Table \ref{table:1}.

\subsubsection{Empirical significance levels}

Figure \ref{fig:Sig_Heavy} shows the empirical significance levels. 
It is evident that the test based on the FL+ permutation is extremely liberal in the presence of continuous nuisance covariates with location-scale shift or noise nuisance effects.  
The results are similar for the 
FL permutation and hence are omitted to increase the readability of Figure \ref{fig:Sig_Heavy}. 
Moreover, a similar behavior is observed for the PH test for small sample sizes (less than 500). For large sample sizes (more than 500), the overall behaviour of the method is unpredictable. Furthermore, as expected, the NC test is liberal. 
In contrast, the empirical significance levels of the RL, RLS, RQ and WN tests were close to the nominal level, independently of the type of the nuisance effect or the sample sizes.

\begin{figure}
    \centering
    \includegraphics[scale=0.7]{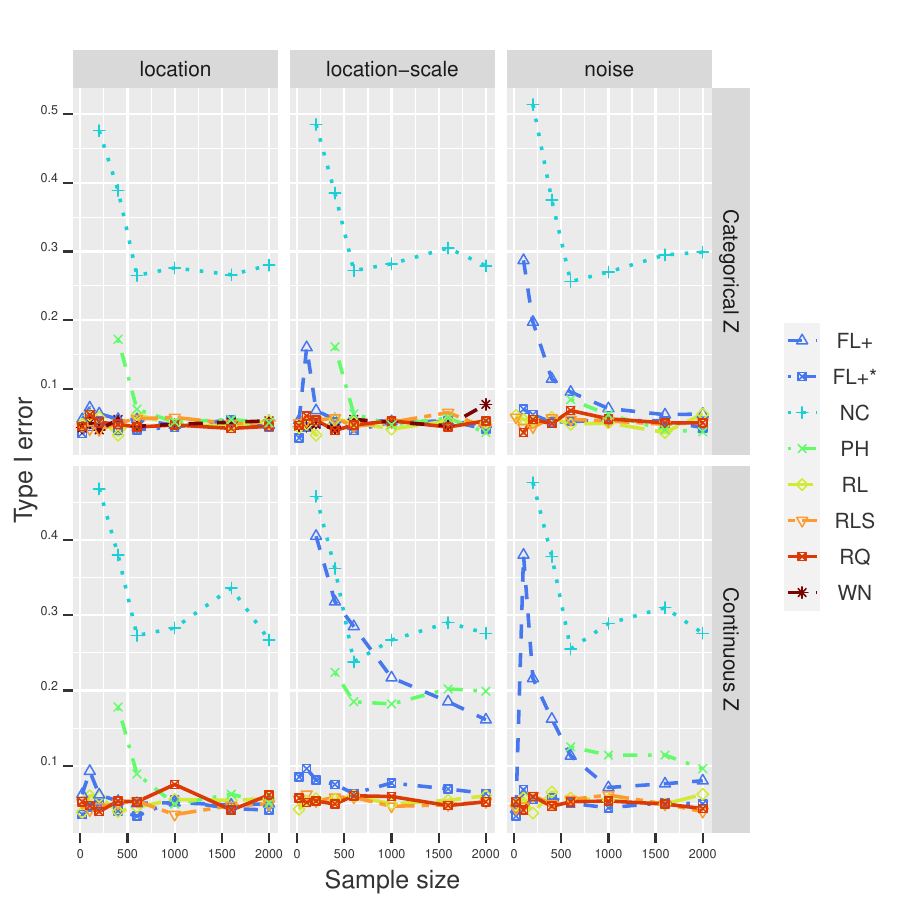}
    \caption{Empirical significance levels for Experiments \eqref{Exp:1} and \eqref{Exp:2} among 1000 simulated samples of different sizes (x-axis) for the different tests of Table \ref{table:1} (different colours). The nuisance covariate $Z$ is either categorical or continuous with location, location-scale (Experiment \eqref{Exp:1}) or noise (Experiment \eqref{Exp:2}) effects on the response. Results are based on 10 values for $\tau$ varying from $0.01$ to $0.99$, except the FL+$^*$ test that considers 10 different values of $\tau$'s in the range [0.1,0.9].}
    \label{fig:Sig_Heavy}
\end{figure}

\subsubsection{Power}
\label{sec:Power Ex I and II}
Next the power of those methods that achieved nominal significance levels was studied (see Figure \ref{fig:Power_heavy_tails}). We investigated the power only for the cases and samples sizes where their empirical significance levels were approximately $5\%$. The results suggest that the global envelope tests (the first six test of Table \ref{table:1}) are generally more powerful than the PH test and the FL$+^*$ method. However, the RQ method was an exception; it had lower power than PH test for sample sizes less than 1000. This is likely because the quantile effect is poorly estimated for extreme quantiles. 
The FL+$^*$ method is naturally less powerful as it does not consider the extreme quantiles ($\tau\in[0.1,0.9]$), and the distributional differences between the two groups in Experiments \eqref{Exp:1} and \eqref{Exp:2} were in the tails.  
In the case of continuous location-scale effect the RLS permutation outperformed the RL permutation. On the contrary, under model misspecification, i.e., noise effect,
the RL permutation outperformed the RLS permutation. 
Finally, for location effects (first column of Figure \ref{fig:Power_heavy_tails}) it is unclear which method is the best as the FL+, RL and RLS methods had equally high power, and also WN was equally powerful in the case of categorical $Z$.

\begin{figure}
    \centering
    \includegraphics[scale=0.7]{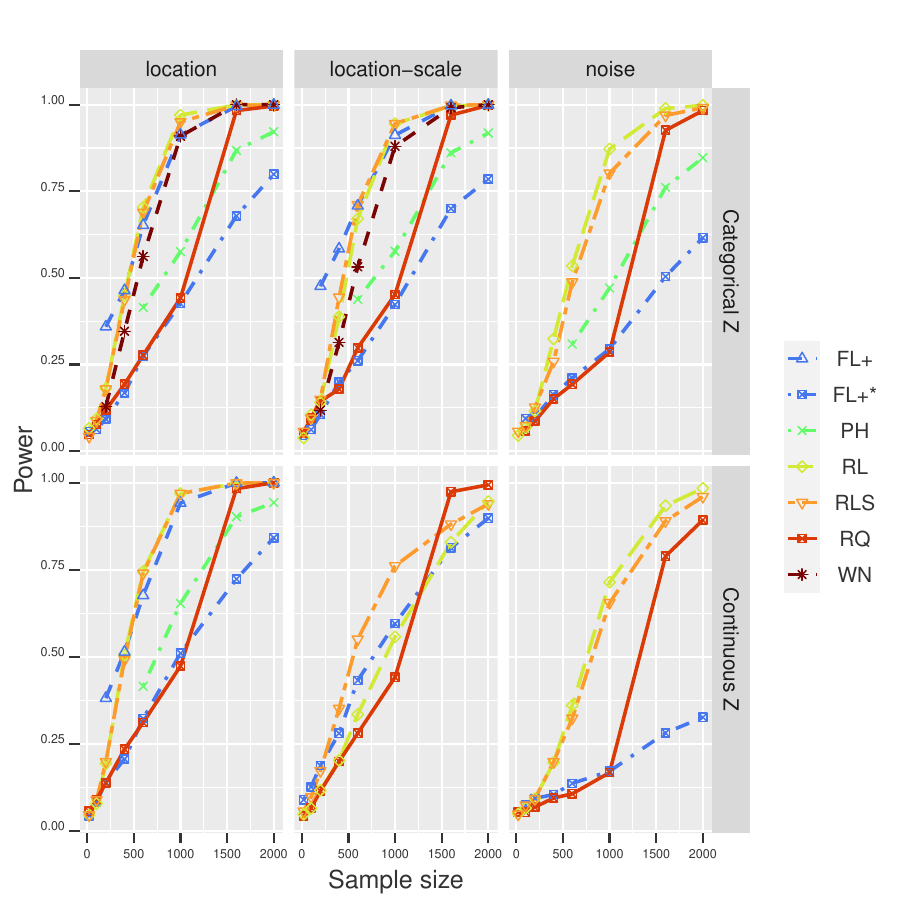}
    \caption{Power for Experiments \eqref{Exp:1} and \eqref{Exp:2} among 1000 simulated samples of different sizes (x-axis) for the different tests of Table \ref{table:1} (different colours). The nuisance covariate $Z$ is either categorical or continuous with location, location-scale (Experiment \eqref{Exp:1}) or noise (Experiment \eqref{Exp:2}) effects on the response. Results are based on 10 values for $\tau$ varying from $0.01$ to $0.99$, except the FL+$^*$ test that considers 10 different values of $\tau$'s in the range [0.1,0.9].}
    \label{fig:Power_heavy_tails}
\end{figure}

\subsubsection{Liberality of Freedman-Lane and pointwise $p$-values}

To investigate the source of liberality in the 
FL, FL+ and PH tests, we performed local tests, i.e., tests for single $\tau$ in the setup of Experiment \eqref{Exp:1}. In each such test, only one quantile $\tau$ is considered, and the behavior of the methods is studied.  The individual quantiles considered here were $\tau = 0.01, 0.05, 0.1, 0.2, 0.5$. For a categorical nuisance covariate, the resulting significance levels are shown in Figure \ref{fig:singletau_cat} and for a continuous nuisance covariate the corresponding results are displayed in Figure \ref{fig:single_tau_cont}.
The tests based on the FL and FL+ permutations were extremely liberal for extreme quantiles and the pointwise test was liberal for extreme quantiles and small samples sizes. In the case of continuous nuisance with location-scale effects, the liberality was more apparent. 
On the other hand,
the test based on the FL and FL+ permutations achieved correct significance levels for non-extreme quantiles and hence they are suitable for global testing when quantile range excludes the most extreme quantiles. For instance, in the case of median regression the use of the FL and FL+ permutations can be justified.
Also for sample sizes larger than 500, it seems acceptable to exclude only quantiles $< 0.1$.

\begin{figure}
    \centering
    \includegraphics[scale=0.7]{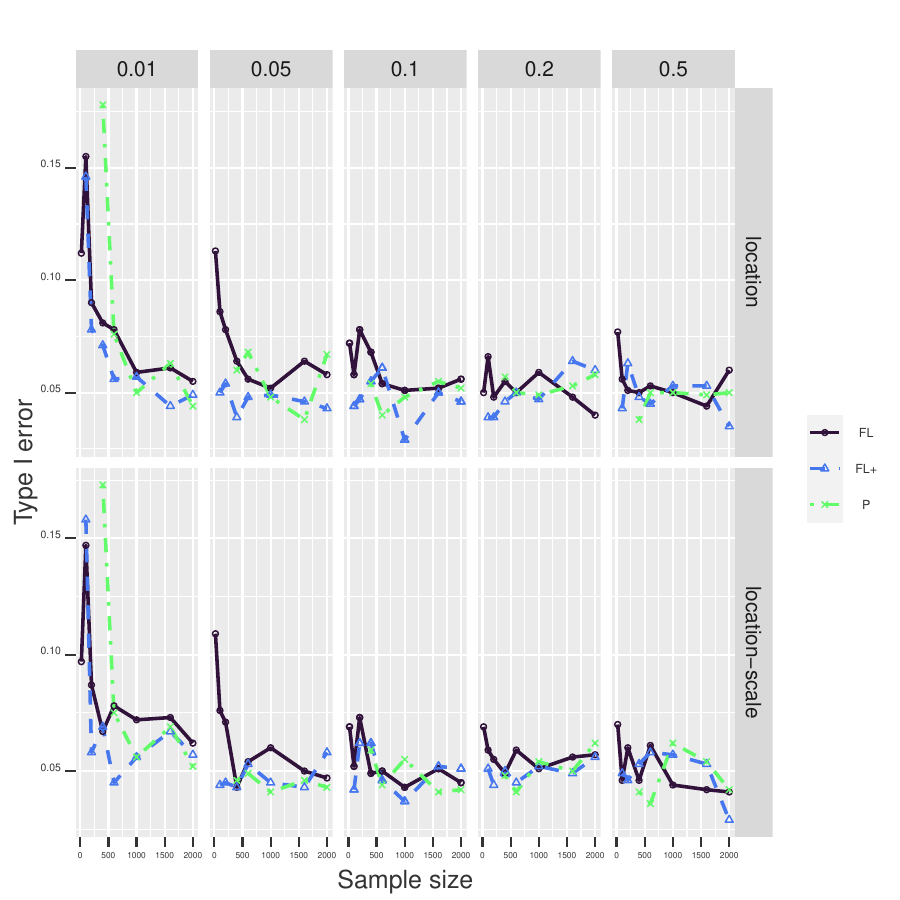}
    \caption{Empirical significance levels for the simulation Experiment \eqref{Exp:1} among 1000 simulated samples of different sizes (x-axis) for the Freedman-Lane based permutation strategies and pointwise $p$-value (different colours). The nuisance covariate $Z$ is categorical  with location or location-scale effects on the response. Quantiles considered are $\tau=\{0.01,0.05,0.1,0.2,0.5\}$ (columns).}
    \label{fig:singletau_cat}
\end{figure}

\begin{figure}
    \centering
    \includegraphics[scale=0.7]{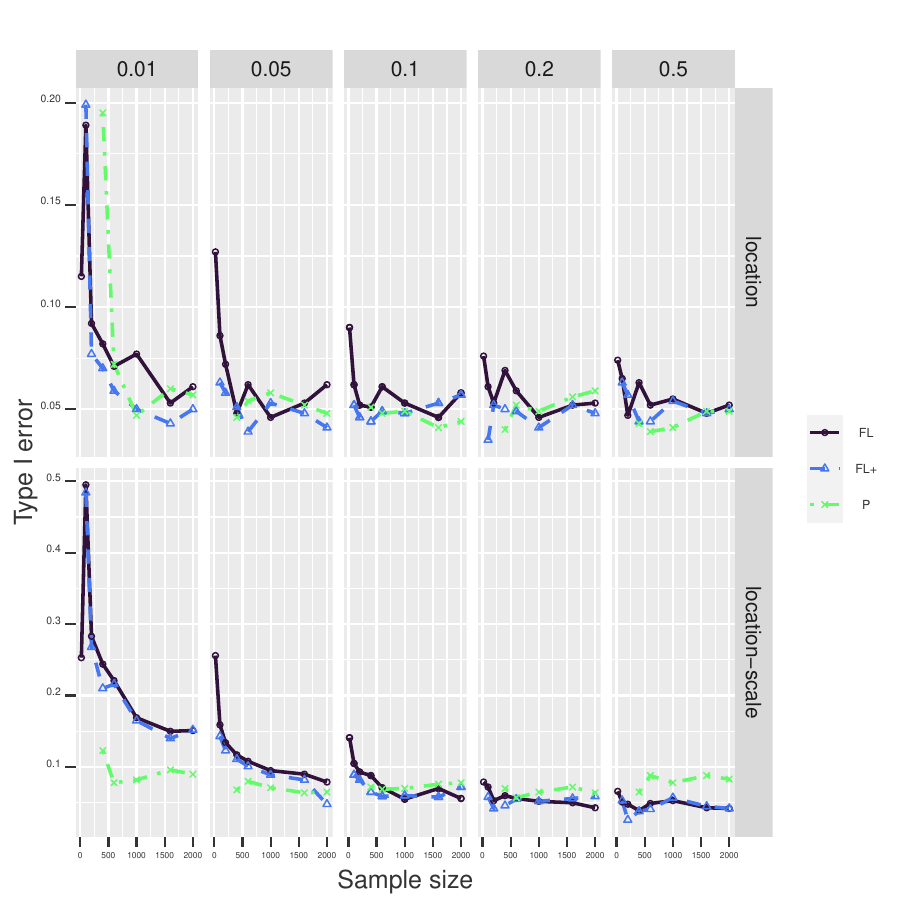}
    \caption{Empirical significance level for the simulation Experiment \eqref{Exp:1} among 1000 simulated samples of different sizes (x-axis) for the Freedman-Lane based permutation strategies and pointwise $p$-value (different colours). The nuisance covariate $Z$ is continuous  with location or location-scale  effects on the response. Quantiles considered are $\tau=\{0.01,0.05,0.1,0.2,0.5\}$ (columns).}
    \label{fig:single_tau_cont}
\end{figure}

\subsection{Sensitivity to effects on the scale of the distribution}

The performance of the methods was studied in two further cases where $X$ was still categorical, but it affected the scale of the response distribution. The conditional response distribution was defined through a $t_{df}$ distribution where the degrees of freedom $df$ were controlled by the realisations of $X$. As the normal distribution coincides with the $t_{df}$ distribution as $df\rightarrow\infty$, the contrast between the standard normal and the $t_4$ distribution (as studied in Section \ref{sec:simstudy_tails}) is larger than the contrast between $t_{df}$ distributions with $df$ simulated from a Poisson distribution with mean 3. Similarly to previous experiments, location, location-scale and noise effects were added to the response distribution. 
In Experiment \eqref{Exp:3},
  \begin{equation}\tag{III}
 \qquad 
 \label{Exp:3}
\left\{
	\begin{array}{ll}
		X \sim \max(\text{Poisson}(3),1)\\
   Y'\mid X \sim t_{X}  \\
   Z \sim F_Z\\
   Y =(1+aZ)Y'+bZ
	\end{array}
\right.\end{equation}
As before, $a,b\in \mathbb{R}$ are parameters controlling the size of the nuisance effect and $F_Z$ is the distribution of the nuisance variable, for which we considered the the same cases (Ia)-(Id) as in Experiment \eqref{Exp:1}. In Experiment \eqref{Exp:4},

\begin{equation}
\tag{IV}\label{Exp:4}
 \qquad 
\left\{
	\begin{array}{ll}
		X \sim \max(\text{Poisson}(3),1)\\
   Y'\mid X \sim t_{X}  \\
   
   Z \sim F_{Z} \\
   Z_1 \sim \text{Unif}(0,1.5)\\ \epsilon \sim N(1,0.04)\\
   Y = \left\{
	\begin{array}{ll}
 \epsilon & \text{if } Z < Z_1
 \\
   Y' & \text{otherwise}
 \end{array}
 \right.
	\end{array}
\right.
	\end{equation}
where $Z$ and $Z_1$ are nuisance covariates and $F_Z$ is the distribution of the nuisance covariate $Z$ with the two cases (IIa)-(IIb) as in Experiment \eqref{Exp:2}.   

As shown in Figure \ref{fig:student_significance}, the FL+ and PH tests, are again liberal when extreme quantiles are considered. 
Regarding the significance levels the other tests also behaved similarly as in the previous experiments: the NC method 
was highly liberal and PH, RL, RLS, RQ and WN were fine. 

  \begin{figure}
    \centering
    \includegraphics[scale=0.7]{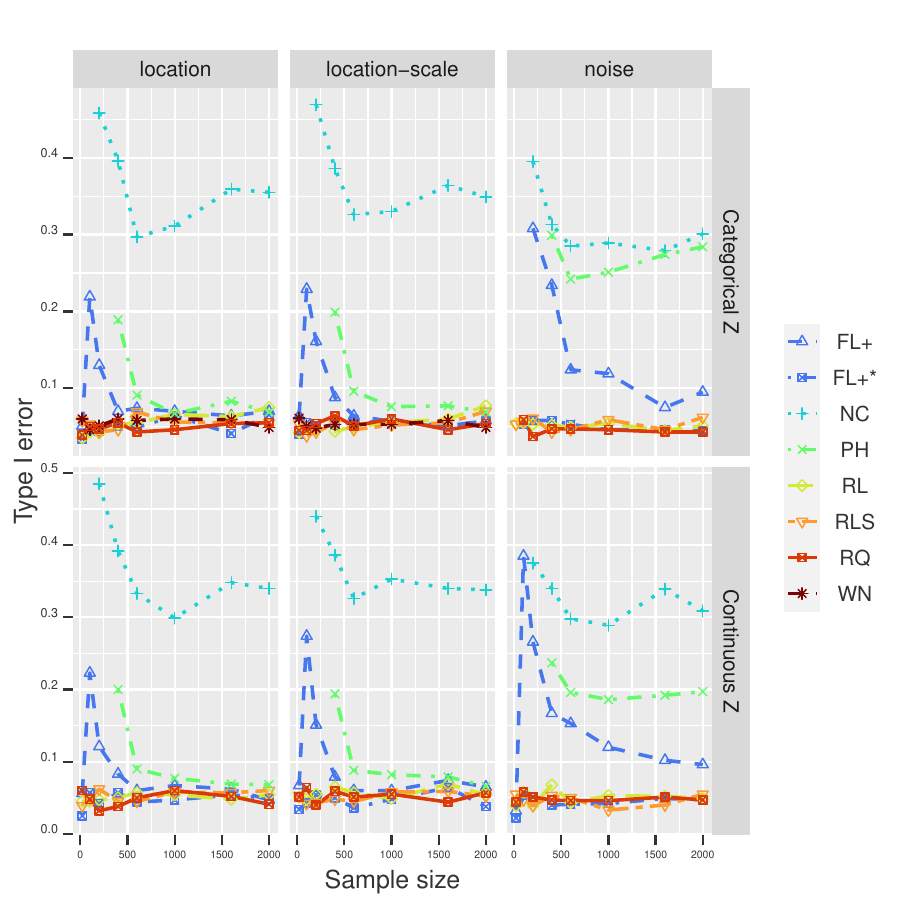}
    \caption{Empirical significance levels for Experiments \eqref{Exp:3} and \eqref{Exp:4} among 1000 simulated samples of different sizes (x-axis) for the different tests of Table \ref{table:1} (different colours). The nuisance covariate $Z$ is either categorical or continuous with location, location-scale (Experiment \eqref{Exp:3}) or noise (Experiment \eqref{Exp:4}) effects on the response. Results are based on 10 values for $\tau$ varying from $0.01$ to $0.99$, except the FL+$^*$ test that considers 10 different values of $\tau$'s in the range [0.1,0.9].}
    \label{fig:student_significance}
\end{figure}

The power of the methods was also investigated (see Figure \ref{fig:student_power}). As in Section \ref{sec:Power Ex I and II}, we only considered the samples sizes and methods with a significance level of approximately 5\%. In the presence of location and location-scale effects the RQ permutation had the highest power with the RL and RLS being the less powerful methods.
\begin{figure}
    \centering
    \includegraphics[scale=0.7]{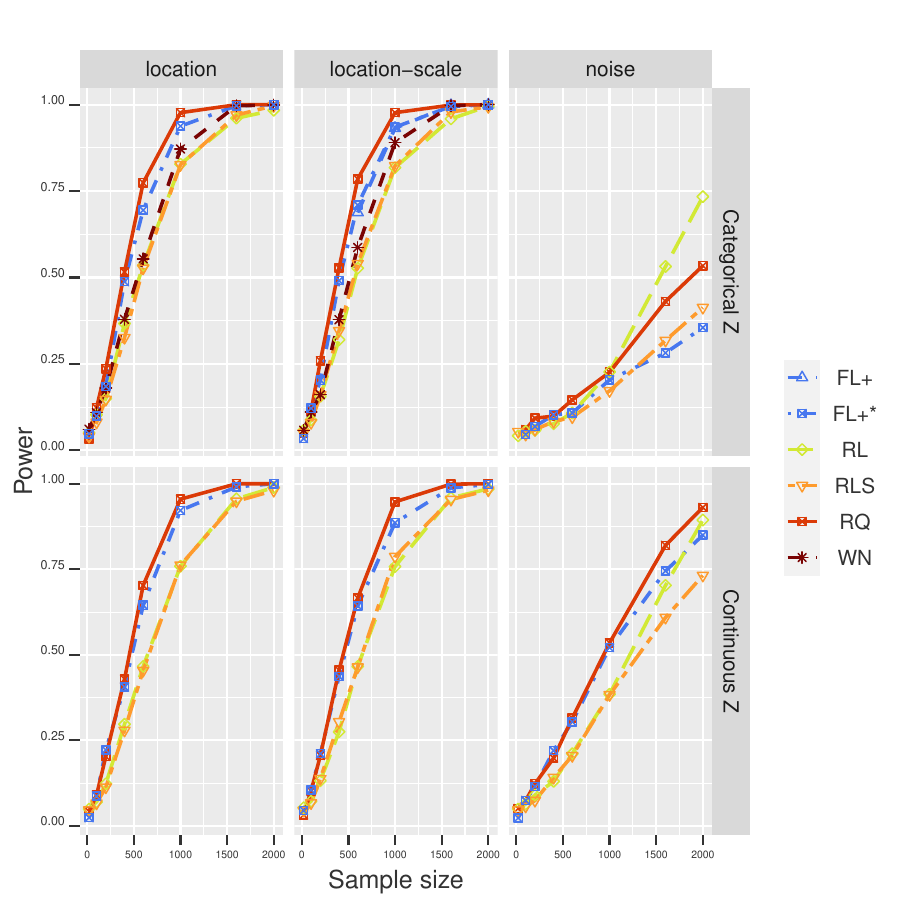}
    \caption{Power for Experiments \eqref{Exp:3} and \eqref{Exp:4} among 1000 simulated samples of different sizes (x-axis) for the different tests of Table \ref{table:1} (different colours). The nuisance covariate $Z$ is either categorical or continuous with location, location-scale (Experiment \eqref{Exp:3}) or noise effects (Experiment \eqref{Exp:4}) on the response. Results are based on 10 values for $\tau$ varying from $0.01$ to $0.99$, except the FL+$^*$ test that considers 10 different values of $\tau$'s in the range [0.1,0.9].}
    \label{fig:student_power}
\end{figure}

\subsection{Sensitivity to effects on the shape of the distribution}

Finally, we considered the case where $X$ is either discrete or continuous and influences the shape of the response distribution while the nuisance covariate $Z$ influences the scale of the response distribution. This Experiment \eqref{Exp:5} is in detail as follows:
\begin{equation}\tag{V}
 \qquad  
\left\{
	\begin{array}{ll}
		X \sim F_X\\
  Z\sim \text{Unif}(0.5,2)\\
  Y\sim  \text{Gamma}(X,Z)\label{Exp:5}
	\end{array}
\right.\end{equation}
where $F_X = \text{Unif}(4,5)$ if $X$ is continuous and $F_X$ takes values $4.7$ and $5$ with equal probabilities if $X$ is categorical.

We studied at the empirical significance levels in this experiment by simulating the interesting covariate $X$ having no effect on the response distribution, i.e., the data ($Y$) were simulated from the Gamma distribution with shape parameter $shape=4.5$. Again
the FL+
and the PH tests were liberal when extreme quantiles $\tau$ were considered, while the 
tests with the RL, RLS and RQ permutations achieved correct significance level for all sample sizes (Figure \ref{fig:gamma_sig}).

For testing the power of the tests, the response variable $Y$ was simulated from a Gamma distribution where the shape parameter was defined through the interesting covariate $X$ as specified in \eqref{Exp:5}. As earlier, we considered only the sample sizes and methods whose empirical significance levels were approximately 5\%.
Figure \ref{fig:gamma_pow} shows the results.
The RQ test had low power for small samples, while the other methods were equivalent in terms of power.  

\begin{figure}
    \centering
    \includegraphics[scale=0.7]{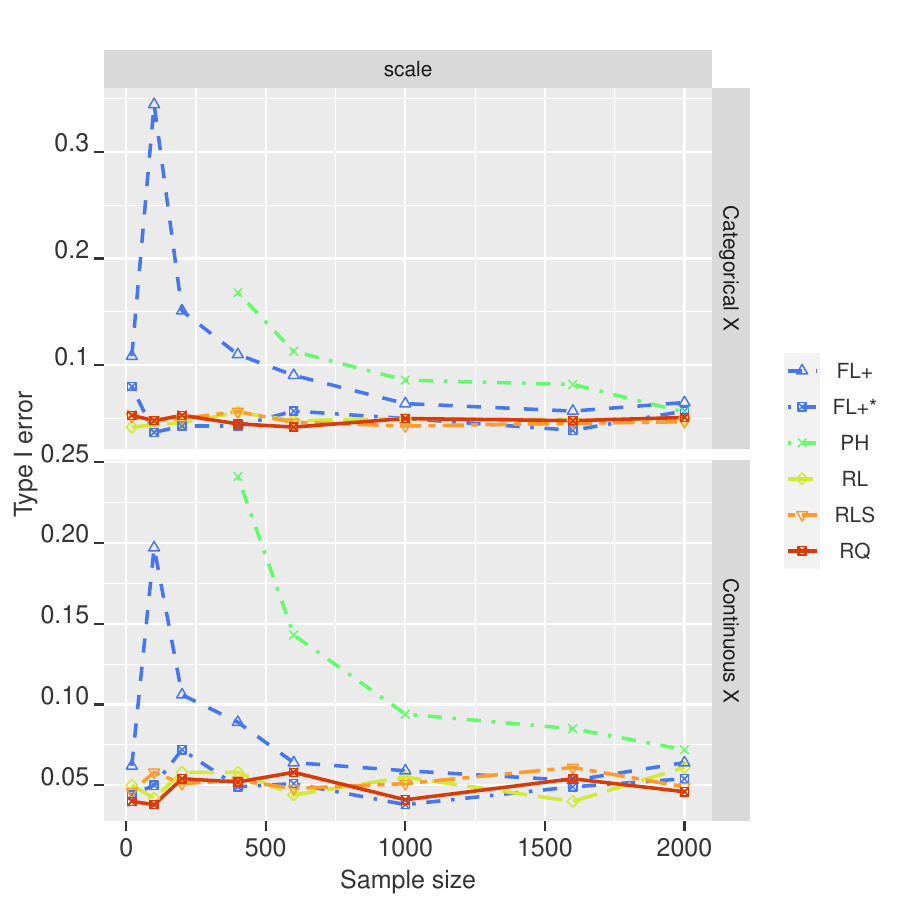}
    \caption{Empirical significance level for Experiment \eqref{Exp:5} among 1000 simulated samples of different sizes (x-axis) for the different tests of Table \ref{table:1} (different colours). The interesting covariate $X$ is either categorical or continuous and the nuisance covariate $Z$ is  continuous with scale effect. Results are based on 10 values for $\tau$ varying from $0.01$ to $0.99$, except the FL+$^*$ test that considers 10 different values of $\tau$'s in the range [0.1,0.9].}
    \label{fig:gamma_sig}
\end{figure}
\begin{figure}
    \centering
    \includegraphics[scale=0.7]{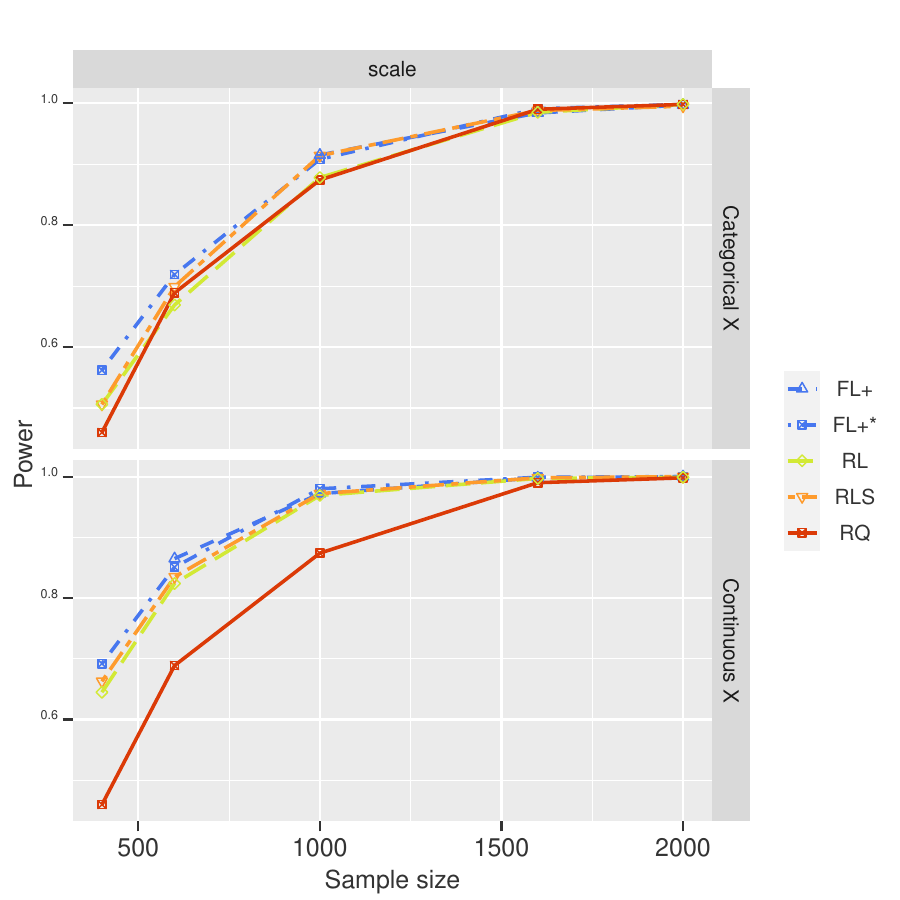}
    \caption{Power for Experiment \eqref{Exp:5} among 1000 simulated samplesof different sizes (x-axis) for the different tests of Table \ref{table:1} (different colours). The interesting covariate $X$ is either categorical or continuous and the nuisance covariate $Z$ is  continuous with scale effect. Results are based on 10 values for $\tau$ varying from $0.01$ to $0.99$, except the FL+$^*$ test that considers 10 different values of $\tau$'s in the range [0.1,0.9].}
    \label{fig:gamma_pow}
\end{figure}

\subsection{Sensitivity to correlation of the interesting and nuisance covariates}

Finally we studied the performance of the permutation methods in the case where the interesting covariate $X$ and the nuisance covariate $Z$ are correlated. In this Experiment \eqref{Exp:6}, we had

\begin{equation}\tag{VI}
 \qquad  
\left\{
	\begin{array}{ll}
		A,B,C \sim \text{Unif}(0,1)\\
  X = (1-c)A+cC , \text{ with } 0\leq c\leq1\\
  Z = (1-c)B+cC, \text{ with } 0\leq c\leq1\\
  Y
  \sim \text{Gamma}(4+X,1+Z)\label{Exp:6}
	\end{array}
\right.\end{equation}

We considered the cases with $c = 0,  0.3, 0.5, 0.7$.  In this setup $X$ and $Z$ are positively correlated with correlation given by $\text{cor}(X,Z) = \frac{c^2}{1+2c^2-2c}$. Therefore, increasing $c$ towards 1, increases the correlation between $X$ and $Z$, while $X$ and $Z$ are independent when $c=0$. To increase the readability of Figure \ref{fig:exp7_significane} showing the results, only Type I errors lower than 0.3 are shown. For instance, under this model misspecification, the RL permutation strategy led to the more liberal test the larger the correlation between $X$ and $Z$ was.  This is because the RL permutation fails to remove the complete nuisance effect, here a location-scale effect, from the response $Y$. Hence, there is still a significant effect of the nuisance $Z$ present on the residuals $\Bepsilon_Z$. Now, as the correlation between $X$ and $Z$ increases, the effect of $X$ on $\Bepsilon_Z$ becomes significant causing the test to be more liberal. 
On the contrary, the permutation tests that correctly remove the nuisance effects (RLS and RQ) were conservative with increasing correlation, resulting in low power. Finally, only the extension of the Freedman-Lane test without considering extreme quantiles (FL+$^*)$ achieved the significance level close to the nominal level for all levels of correlation.

\begin{figure}[h]
    \centering
    \includegraphics[scale=0.7]{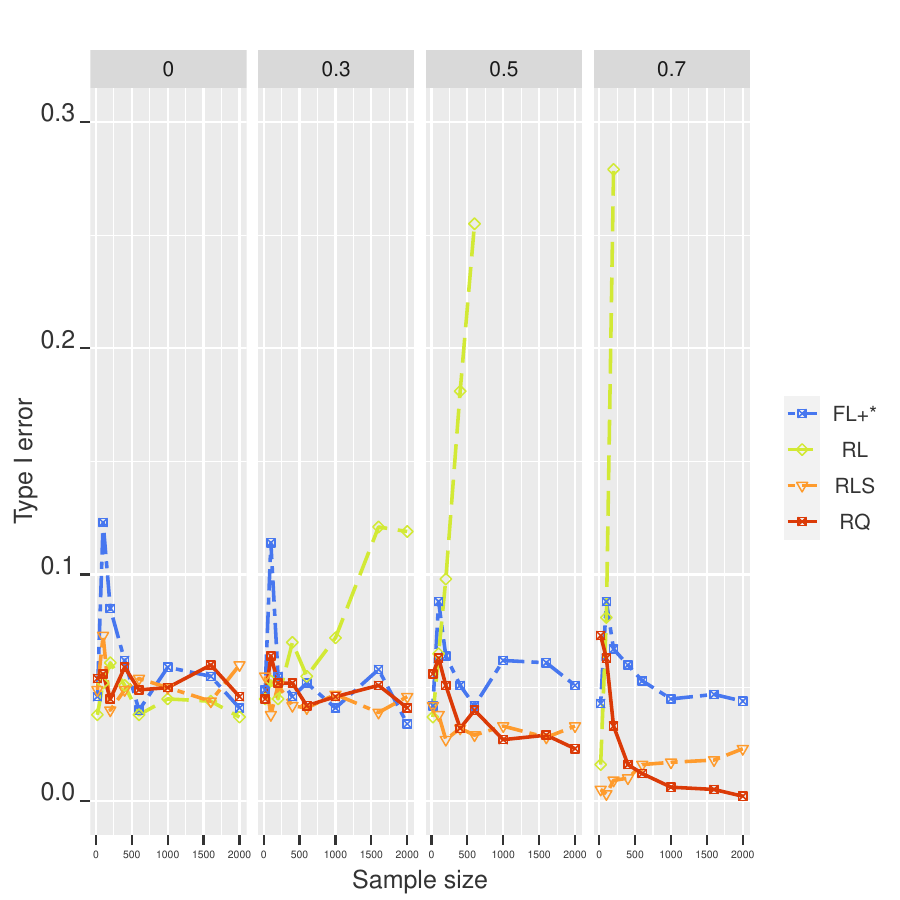}
    \caption{Empirical significance levels for Experiment \eqref{Exp:6} among 1000 simulated samples of different sizes (x-axis) for the different tests of Table \ref{table:1} (different colours). The nuisance covariate $Z$ is continuous with scale effect. The different columns correspond to the results for different values of c. Results are based on 10 values for $\tau$ varying from $0.01$ to $0.99$, except the FL+$^*$ test that considers 10 different values of $\tau$'s in the range [0.1,0.9].}
    \label{fig:exp7_significane}
\end{figure}

Further we studied, the behaviour of the methods in the simulation setup of the first two experiments modified to the case where $X$ and $Z$ are correlated. These two cases are in detail as follows:
\begin{equation}\tag{VII}
 \qquad  
\left\{
	\begin{array}{ll}
		A,B,C \sim \text{Unif}(0,1)\\
  X = \text{round}((1-c)A+cC) , \text{ with } 0\leq c\leq1\\
  Y' \mid X \sim \left\{
	\begin{array}{ll}
 N(0,1) & \text{if } X=0\\
t_4 & \text{if } X=1
 \end{array}
 \right.\\
Z = 1.5\cdot ((1-c)B+cC), \text{ with } 0\leq c\leq1\\
   Y =(1+aZ)Y'+bZ
  \label{Exp:7}
	\end{array}
\right.\end{equation}
\begin{equation}\tag{VIII}
 \qquad  
\left\{
	\begin{array}{ll}
		A,B,C \sim \text{Unif}(0,1)\\
  X = \text{round}((1-c)A+cC) , \text{ with } 0\leq c\leq1\\
    Y' \mid X \sim \left\{
	\begin{array}{ll}
 N(0,1) & \text{if } X=0\\
t_4 & \text{if } X=1
 \end{array}
 \right.\\
  Z = 1.5\cdot ((1-c)B+cC), \text{ with } 0\leq c\leq1\\
    Z_1 \sim \text{Unif}(0,1.5)\\
		 \epsilon \sim N(1,\sigma_\epsilon^2)\\
   Y = \left\{
	\begin{array}{ll}
 \epsilon & \text{if } Z < Z_1
 \\
   Y' & \text{otherwise}
   
  \label{Exp:8}
	 \end{array}
 \right.
	\end{array}
\right.\end{equation}
We considered values $c=0.3, 0.4, 0.5, 0.9$. As before increasing $c$ towards 1 increases the dependency between $X$ and $Z$.

\begin{figure}[h]
    \centering
    \includegraphics[scale=0.7]{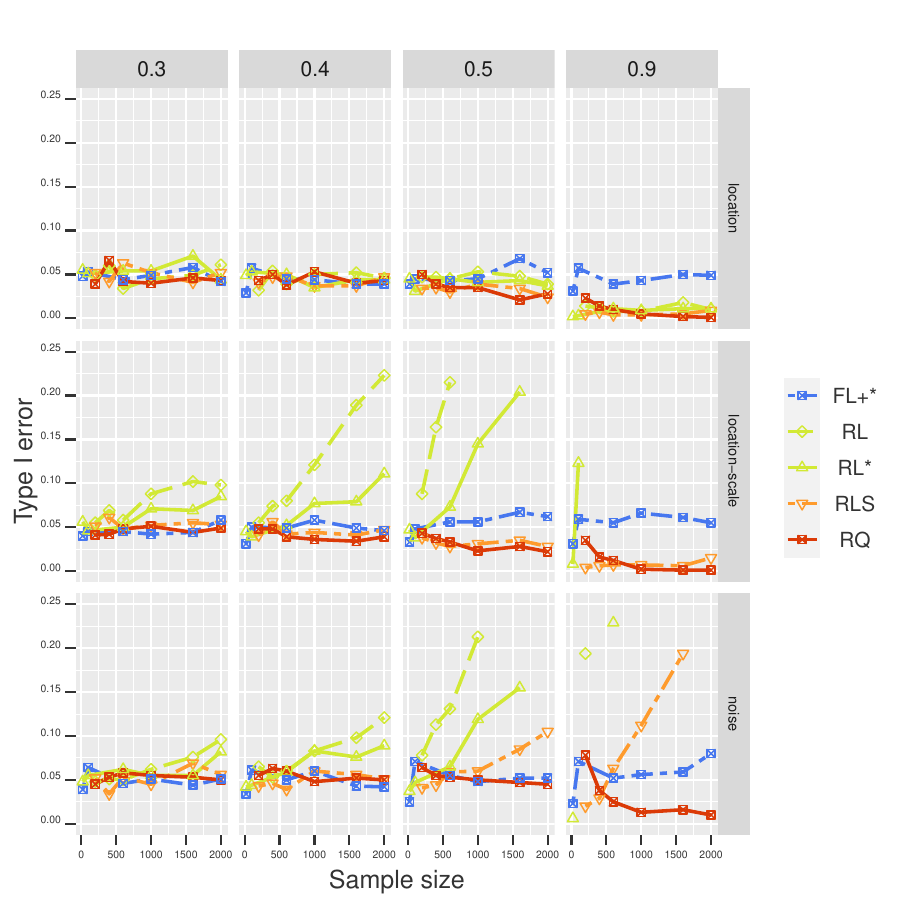}
    \caption{Empirical significance levels for Experiments \eqref{Exp:7} and \eqref{Exp:8} among 1000 simulated samples of different sizes (x-axis) for the different tests of Table \ref{table:1} (different colours). The nuisance covariate $Z$ is continuous with location, location-scale (Experiment \eqref{Exp:7}) or noise (Experiment \eqref{Exp:8}) effects on the response. The different
columns correspond to the results for different values of c. Results are based on 10 values for $\tau$ varying from $0.01$ to $0.99$, except the FL+$^*$ test that considers 10 different values of $\tau$'s in the range [0.1,0.9].}
    \label{fig:heavy_corr}
\end{figure}
The empirical significance levels for Experiments \eqref{Exp:7} and \eqref{Exp:8} are shown in Figure \ref{fig:heavy_corr}.
According to the results, the RL permutation strategy is liberal when the assumption of the test, i.e., the effect of nuisance is only in location, is not satisfied (cases of location-scale and noise). That is caused by the fact that RL method filters away only the location effect of nuisance, i.e., the residuals $\Bepsilon_{Z}$ still contain other effects of the nuisance. As a result, if the interesting and the nuisance covariates are correlated, the interesting covariate also affects $\Bepsilon_Z$. This remaining effect causes a significant result when the interesting covariate is tested by quantile regression.
The same can be seen for RLS permutation strategy when the assumption of the test, i.e., the effect of nuisance is only in location and scale, is not satisfied. This effect is not presented in RQ permutation strategy, nevertheless all three methods appear to be conservative with increasing correlation between interesting and nuisance covariates. 

The results of Experiments \eqref{Exp:6} and \eqref{Exp:7} suggest that the effect of nuisance covariates must be rigorously tested. This is possible via Khmaladze test implemented in the quantreg package. Also the visual inspection of the pointwise confidence bands provided by quantreg package is useful, since the Khmaladze test is recommended for non extreme quantiles only. On the other hand, if the interest is not on the tails of the distribution, then the FL+ permutation test without extreme quantiles appears to be a good choice: it achieved the correct significance level independently of the amount of correlation between $X$ and $Z$ in our experiments.

\subsection{Summary}

From the above experiments we give the following recommendations: 
\begin{itemize}
    \item The pointwise minimum $p$-value is extremely liberal for the global test as the multiple testing problem is not considered.
    \item The Freedman-Lane based global quantile tests should be avoided when extreme quantiles are considered. If the interval for quantiles is (0.1,0.9), then the Freedman-Lane global quantile tests should be avoided with less than 500 data.   
    \item The  PH test seems to be liberal even thought the Holm-Bonferroni correction for multiple testing is conservative.
    \item In the presence of only one categorical nuisance, the WN method is recommended. 
    \item  The RL, RLS methods are liberal when $X$ and $Z$ are correlated, if the assumptions of the effect of nuisance covariates on data are not satisfied.  
    \item The RQ permutation can have  lower power for small sample sizes as the quantile effect is badly estimated for extreme quantiles.
    \item If the nuisance influences only the location, then the RL permutation is recommended and if it further influences the scale then the RLS permutation is recommended. If the effect is unknown, then the RQ permutation is recommended. If tails are not of interest, then the Freedman-Lane without extreme quantiles can be also used.
 
\end{itemize}

\section{Data examples}\label{sec:dataexamples}

\subsection{Forest stand age with respect to forest naturalness}

In the Finnish national forest inventory (NFI), naturalness of the forest is evaluated in the field from three criteria, namely structure, deadwood and human action. \citet{MyllymakiEtal2023} studied the properties of the forest structure within the structural naturalness, and we are also inspecting only this structural naturalness here.
Namely, we investigated the distributions of stand age in the three naturalness groups 'natural', 'near-natural' and 'non-natural' in the Finnish Lapland, excluding the northernmost part. The study region corresponds to 'North' of Myllymäki et al. (2022, Figure 1).
Here, for simplicity, we restricted our attention to plots on rich mineral soils. 
Because the stand age depends potentially on the dominant species, we included as the nuisance covariate the dominant species as a variable with three categories 'Broadleaf', 'Conifer' and 'Mixed' as defined in Myllymäki et al. (2022). Numbers of plots in each category are shown in Table \ref{tab:NFIplots}.

Our quantile regression model is 
$$
Age \sim constant + naturalness + species
$$
where naturalness is our interesting factor and species is the nuisance.
According to the quantile regression fit (Figure \ref{fig:ex_NFI}, rows 1-3), the effect of dominant species appears to be location-scale shift{, since the estimated coefficients (rows 1 and 2) appear to be linear in $\tau$.} 
Therefore, to test for the differences between the distributions of stand age in the natural, near-natural and non-natural forests, we applied the permutation algorithm RLS of Section \ref{sec:cont_permstrategy2}. 
Figure \ref{fig:ex_NFI} (row 4) shows the results of this test based on 2499 permutations and $\tau_i = 0.051 + (0.949-0.51)i/99$ for $i=0,1, \dots, 99$.
The global envelope is shown by grey zone, while the estimated coefficients are shown by black solid line, overlaid with red dots when outside the envelope. 
Note here that the global test of naturalness contains both functional coefficients shown in row 4 of Figure \ref{fig:ex_NFI}, thus the test corresponds to the ANOVA test of the effect of the categorical covariate, which is tested using $2\times 99$  pointwise tests. Thus the test identifies both the significant quantiles and the corresponding coefficient which are significant under the global test. 
Here the coefficients of near-natural and natural forests show the difference to non-natural reference group. 
It can be seen that both the near-natural and natural forest are uniformly older than non-natural forests for all quantiles. 

For another example we switched the roles of naturalness and dominant species. 
Since the effect of naturalness on stand age appears not to be a location-scale shift{, since the naturalness coefficients in rows 2 and 3 do not appear to be linear}, we used the RQ permutation strategy. 
We used again 2499 permutations and the same $\tau$s as earlier. 
Figure \ref{fig:ex_NFI} (row 4) shows the results of this global test.
It can be seen that there is a significant effect for quantiles between 0.3 and 0.85.
This means that the stand age distribution of broadleaf dominated forests is more skewed to the left than the distribution of conifer dominated forests, but the ranges are equal.
The mixed forests are also younger than the conifer dominated forests for some quantiles between 0.65 and 0.8, suggesting that the difference is present only for older stands.

\begin{table}[ht!]
    \centering
    \begin{tabular}{c|cccc}
Dominant species &  0 & 1 & 2 \\ \hline
Broadleaf &	 30	& 9	 & 81 \\
Conifer	  &  59  & 36 & 	342 \\
Mixed	 &  54  & 23 & 	139	\\
    \end{tabular}
    \caption{Numbers of NFI plots in total and in the different naturalness groups (0 = natural; 1 = near-natural; 2 = non-natural).}
    \label{tab:NFIplots}
\end{table}

\begin{figure}
    \centering
    \includegraphics[width=13cm]{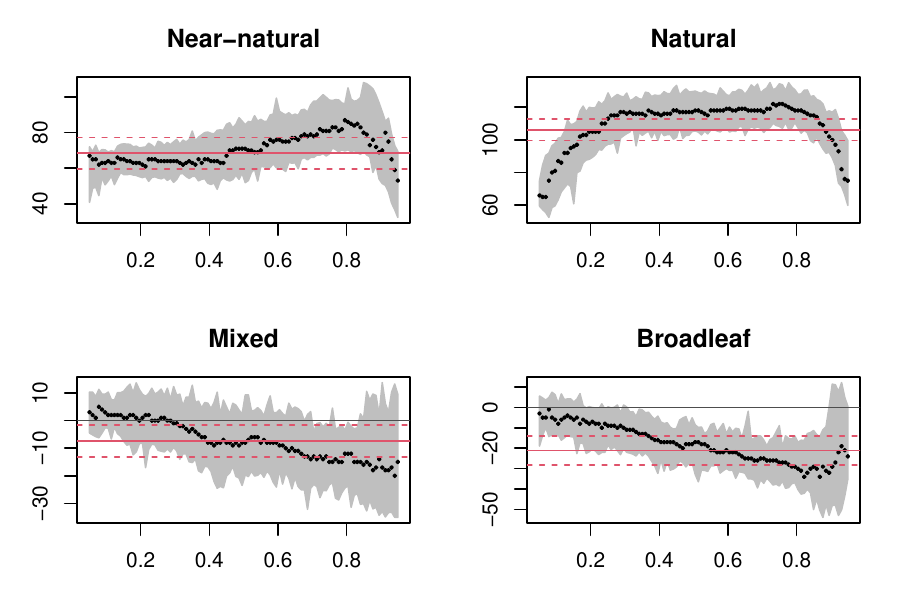}
    \includegraphics[width=13cm]{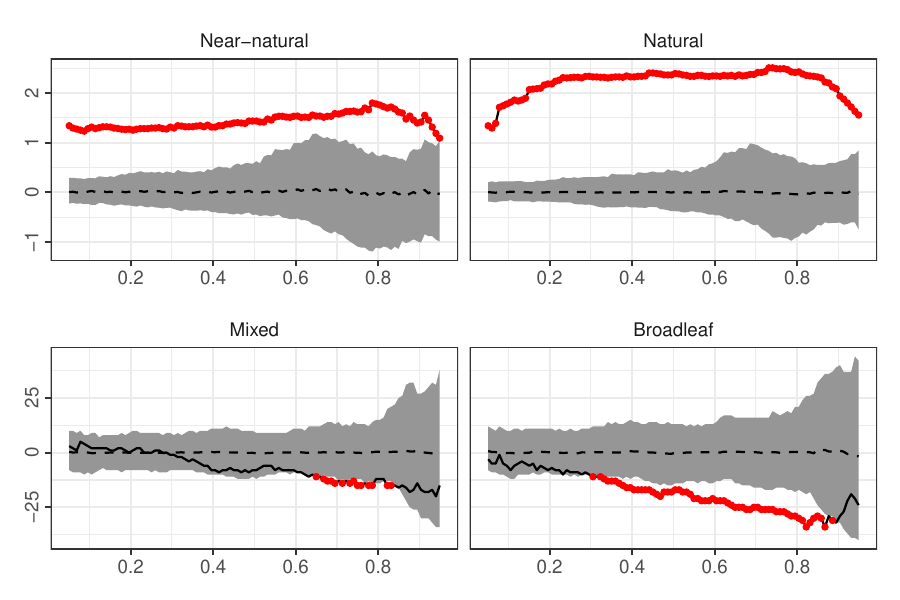}
    \caption{95\% pointwise confidence bands (rows 1-2) and 95\% global envelopes (rows 3 and 4) for the effect of naturalness or dominant species on stand age. The global envelope on row 3 is based on the RLS permutation strategy testing the effect of naturalness accounting for the dominant species as a nuisance.
    Row 4 is based on the RQ permutation strategy testing the effect of dominant species accounting for the naturalness as nuisance. }
    \label{fig:ex_NFI}
\end{figure}

\subsection{Effect of gold on log returns for exchange courses}

Here we investigated the effect of price of gold on the log return for exchange rates.
For the sake of this example, we think that the log returns of exchange rates might also be affected by the prices of oil and uranium.
To remove the effect of inflation from the prices, the prices of gold, oil and uranium were computed as residuals of a simple exponential model that was fitted to the original prices.  The data contain 3201 observations.

The top row of Figure \ref{fig:example_gold} shows the result of quantile regression with pointwise confidence bands. The pointwise confidence bands suggests the presence of the effect of gold on the log returns.
To account for the multiple testing problem we applied the global test with RLS permutation strategy.
The RLS permutation strategy was chosen because the prices of uranium and oil appear to be location scale shifts; their coefficients behave quite linearly with respect to $\tau$. This can not be said about the gold coefficients, which justifies the quantile regression approach.
The model included the prices of oil and uranium as nuisance covariates. 
The result of the global quantile regression test is shown in the third row, first column of Figure \ref{fig:example_gold}. Since the estimated coefficients of gold do not leave the global envelope in any point, we can not reject the hypothesis of no influence of log returns by gold. This result is accompanied by the $p$-value=0.48.

Switching the roles of nuisance and interesting covariates, we can observe the effect of oil and uranium from the results of global quantile regression (Figure \ref{fig:example_gold} second row, second and third column). In these two tests the RQ permutation strategy was used for security since the effect of gold seems to be non-linear. The results show that the increase of oil price significantly reduces volatility of log returns. (For low quantiles, the coefficient of oil is significantly positive, and for high quantiles it is negative.) 
The increase of uranium prices decreases the regression coefficients significantly only for low quantiles, meaning the increased possibility for a big fall of the exchange course.

\begin{figure}
    \centering
    \includegraphics[width=\textwidth]{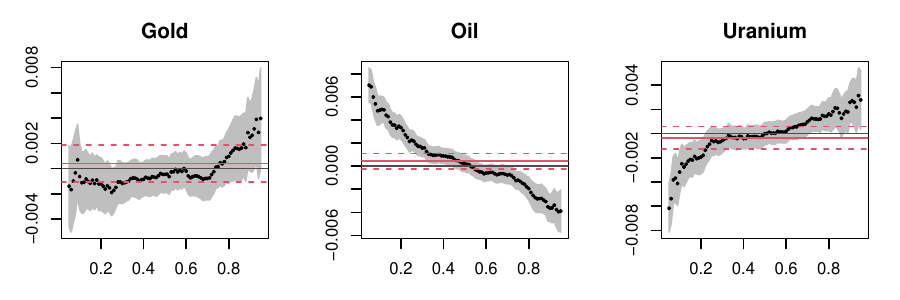}
    \includegraphics[width=\textwidth]{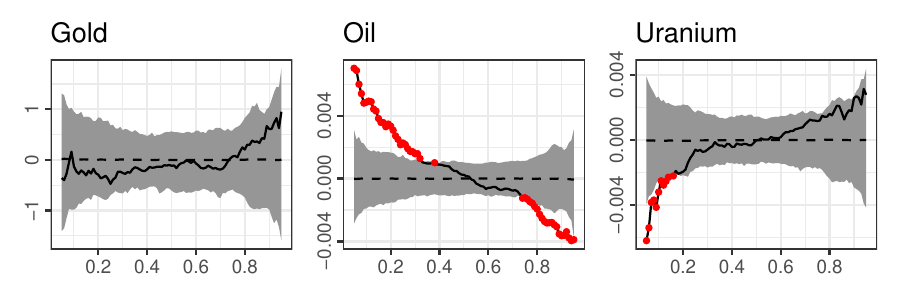}
    \caption{95\% pointwise confidence bands (top row) and 95\% global envelopes (bottom row) for price of gold, oil or uranium as the interesting covariate and having all others as nuisances, using 2499 permutations and the RLS, RQ or RQ permutation strategy, repectively.    
    }

    \label{fig:example_gold}
\end{figure}

\section{Conclusions and discussion}\label{sec:discussion}

In this paper we studied the possibilities to test the significance of a covariate in global quantile regression, i.e., simultaneously for all the quantiles. We realized first that the pointwise $p$-values traditionally used in quantile regression are seriously liberal for extreme quantiles, so much that even the conservative multiple testing adjustment (cf.\ PH of Table \ref{table:1}) does not correct the liberality. Therefore we decided to rely on pointwise permutation tests with the global envelope test as the multiple testing adjustment procedure. 

The choice of the permutation strategy is the crucial point in permutation tests. Surprisingly, it turns out that the traditionally used Freedman-Lane permutation strategies are also liberal for extreme quantiles. Therefore, we proposed other permutation strategies which seem to work well even for extreme quantiles. These strategies are based on evaluating the type of influence of data by nuisance covariates. If this influence is only in location, the permutation with removal of the location effect is recommended. If this influence is in location and scale, the permutation with removal of the location and scale effect of nuisance covariates is recommended. If this influence is more general, then the permutation with removal of the quantile effect of the nuisance covariates is recommended. The recommended methods were  conservative when correlation between nuisance and interesting covariate was present and the assumptions of these methods about the effect of nuisance covariate on the data were satisfied. {We believe that this is always the case as if the model is correctly specified $\Bepsilon_Z$ will not contain any nuisance effect, and hence if $X$ and $Z$ are highly correlated $X$ will have no effect on $\Bepsilon_Z$ which will lead to conservative test}. On the other hand, the recommended methods seem to be extremely liberal when the correlation of interesting and nuisance covariates is present and the assumptions of these methods about the effect of nuisance covariate on the data are not satisfied. This behavior makes the assumption of the effect of nuisance covariates on the data critical for choosing the permutation strategy. The reason for that is that the safe method, permutation with removal of the quantile effect of the nuisance covariates, can have lower power than the other proposed methods for smaller amount of data.   

The data study examples show how one can choose the appropriate permutation strategy. They also show that if the pointwise tests are significant, the global test can be significant as well or also must not be.

The proposed tests are useful if we are interested in the existence of the effect of a covariate on the data distribution in at least one quantile. They are also useful if several data distributions are compared but the data are attached with nuisance covariates. An example is that the distribution of a statistic is compared for different health statuses but every person for which the statistic is computed is attached with various covariates like age or sex.

One of the advantages of the global envelope test used on the pointwise permutation tests here is that it provides the graphical output which automatically detects the quantiles responsible for the potential rejection. Also it automatically detects which levels of the categorical covariates differ from the overall mean across all levels. Another advantage of the global envelope test here is its nonparametric nature which causes that the adjustment procedure is valid for any test statistic without necessity of computing its asymptotic variances. 

The only problem in this kind of permutation procedures is the assumption of exchageability of the test vector under the permutation strategy. It is known that when nuisance covariates are present the exchangeability can not be reached even for linear models where the mean value is modelled. For these models the Freedman-Lane procedure is well accepted and the exactness of such tests is studied via simulations. We followed here the same strategy for quantile regression. By our simulation study, we showed that even though our proposed permutation strategies do not reach exchangeability, their empirical significant levels were very close to the nominal level or below it (conservativeness). 
The conservativeness of our procedures appeared only when the nuisance and interesting covariates were correlated.

The proposed procedures were studied only in the cases of main effect models. It is possible to apply our methods also in the case of studying interactions but the proposed permutation strategies would have to be slightly changed, the main effects considered as the nuisance effects would have to appear also in the step 2. of the proposed procedures even thought their effect was already removed in the step 1. This {adjusted} procedure was not rigorously analysed yet and therefore it remains for our future work.

\bibliographystyle{Chicago}


\end{document}